\documentclass[prd,nofootinbib,twocolumn,superscriptaddress,preprintnumbers,balancelastpage,longbibliography]{revtex4-1}
\usepackage{ae,aecompl}
\usepackage{graphicx}	
\usepackage{amsmath}	
\usepackage{graphicx}
\usepackage{dcolumn}
\usepackage{bm}
\usepackage{graphics}
\usepackage{afterpage}
\usepackage{float}
\usepackage{subfigure}
\usepackage{rotating}
\usepackage{multirow}
\usepackage{fancyheadings}
\usepackage{theorem}
\usepackage{moreverb}
\usepackage{euscript}
\usepackage{psfrag}
\usepackage{slashed}
\usepackage{mathtools}
\usepackage[colorlinks=true
,urlcolor=blue
,anchorcolor=blue
,citecolor=blue
,filecolor=blue
,linkcolor=blue
,menucolor=blue
,linktocpage=true
,pdfproducer=medialab
,pdfa=true
]{hyperref}
\usepackage{color,xcolor}

\newcommand{\LCDM}   {$\Lambda$CDM}
\newcommand{\planck}{{\it Planck}}

\newcommand  \be    {\begin{equation}}
\newcommand  \ee    {\end{equation}}
\newcommand{\ba}{\begin{eqnarray}}
\newcommand{\ea}{\end{eqnarray}}

\newcommand{\ukam}{\mu{\rm K}\,{\rm arcmin}}

\begin{document}

\title{Neutrino masses and beyond-$\Lambda$CDM cosmology with LSST and future CMB experiments}
\author{Siddharth Mishra-Sharma}
\affiliation{Department of Physics, Princeton University, Princeton, NJ 08544}
\email{smsharma@princeton.edu}
\author{David Alonso}
\affiliation{Department of Physics, University of Oxford, Keble Road, Oxford, OX1 3RH, UK}
\author{Joanna Dunkley}
\affiliation{Department of Physics, Princeton University, Princeton, NJ 08544}
\affiliation{Department of Astrophysical Sciences, Princeton University, Princeton, NJ 08544}

\preprint{PUPT 2551}
\date{\today}

\begin{abstract}
Cosmological measurements over the next decade will enable us to shed light on the content and evolution of the Universe. Complementary measurements of the Cosmic Microwave Background (CMB) and Baryon Acoustic Oscillations are expected to allow an indirect determination of the sum of neutrino masses, within the framework of the flat $\Lambda$CDM model. However, possible deviations from $\Lambda$CDM such as a non-zero cosmological curvature or a dark energy equation of state with $w\neq -1$ would leave similar imprints on the expansion rate of the Universe and clustering of matter. We show how future CMB measurements can be combined with late-time measurements of galaxy clustering and cosmic shear from the Large Synoptic Survey Telescope to alleviate this degeneracy. Together, they are projected to reduce the uncertainty on the neutrino mass sum to 30 meV within this more general cosmological model. Achieving a 3$\sigma$ measurement of the minimal 60~meV mass sum (or 4$\sigma$ assuming $w=-1$) will require a five-fold improved measurement of the optical depth to reionization, obtainable through a large-scale CMB polarization measurement. 
\end{abstract}

\maketitle

\section{Introduction}
\label{sec:intro}

Cosmological data in the coming decade will be used to tackle fundamental questions about the physical make-up of the Universe. The currently favored  model is \LCDM, describing a flat universe with a cosmological constant, cold dark matter, and a near-zero mass of neutrino particles~\citep{Ade:2015xua}. Upcoming data measuring the evolution of cosmic structures will be used to search for deviations from this model.
The only detectable deviation that we have strong reason to expect is the non-zero neutrino mass, with the total mass known to be at least $\sim60$ meV from oscillation experiments
~\citep{Maltoni:2004ei,GonzalezGarcia:2002dz,Mohapatra:2006gs,Feldman:2013vca}, but departures from a cosmological constant or a flat universe are not theoretically excluded~\citep{Ade:2015xua,Moresco:2016nqq}.

Each departure from \LCDM\ imprints a unique signature on cosmological observables including the growth rate of structure and the expansion rate of the Universe, but at any given cosmic epoch their effects will be partly degenerate.
Previous forecasts have shown how measuring the amplitude of matter fluctuations can lead to a detection of neutrino mass, since a higher mass suppresses structure growth. For example, improving the growth rate measured by CMB lensing data, supplemented with measurements of the baryon acoustic oscillation (BAO) scale, should give an uncertainty on the mass sum of $\sim30$ meV~\citep{Allison:2015qca, Abazajian:2016yjj}\footnote{Or $\sim15$ meV with an improved measurement of the optical depth to reionization, which better determines the primordial amplitude of fluctuations.}. Galaxy clustering and lensing data from the Large Synoptic Survey Telescope (LSST) have been forecast to measure the mass to similar precision~\citep{Font-Ribera:2013rwa}. However,~\citet{Allison:2015qca} found that the mass uncertainty when using CMB lensing can triple when allowing $w$ or geometry to vary, and~\citet{Font-Ribera:2013rwa} found that constraints on the dark energy equation of state from the optical surveys are also significantly degraded when the sum of neutrino masses is varied as a free parameter.

In this paper we examine the issue of how complementary datasets, measuring structure formation over a range of cosmic epochs and with different systematic effects, will be able to distinguish between the various departures from \LCDM. For example, if a non-zero neutrino mass is preferred by the data, will we be able to exclude a time-varying dark energy component, or a non-zero geometry?
We present forecasts considering CMB lensing data to be measured from next generation surveys~\cite[e.g.,][]{Abazajian:2016yjj}, galaxy lensing and galaxy clustering from the Large Synoptic Survey Telescope~\citep{Abell:2009aa}, cross-correlations between these datasets, and BAO distance scales from the DESI experiment~\citep{Aghamousa:2016zmz}. These are combined with measurements of primordial CMB fluctuations. We do not consider the effect of cluster counts calibrated through CMB lensing, which may be able to further alleviate these degeneracies when an extended cosmological parameter set is considered~\citep{Madhavacheril:2017onh}. We also do not explore other cosmological datasets including redshift-space distortions or supernova measurements~\citep{dePutter:2009kn,Das:2013aia,Villaescusa-Navarro:2017mfx}. 

The paper is organized as follows. In Sec.~\ref{sec:numasses} we review the cosmological neutrino mass signal and describe the main physical degeneracies.
In Sec.~\ref{sec:datasets} we describe the datasets, systematic effects and nuisance parameters considered. In Sec.~\ref{sec:results} we present forecasts for $\Sigma m_\nu$, curvature, and a time-varying dark energy equation of state.
We show the impact of improved constraints on the optical depth to reionization, and of BAO measurements, and explore the impact of possible systematic effects and assumptions about the optical data. We conclude in Sec.~\ref{sec:conclusions}.

\section{Physical degeneracies}
\label{sec:numasses}

The  effects of massive neutrinos on the Cosmic Microwave Background (CMB) and Large Scale Structure (LSS) have been extensively studied in the literature~\citep{Allison:2015qca,Ade:2015xua,Dolgov:2002wy,Elgaroy:2004rc,Hannestad:2005gj,Fukugita:2005sb,Lesgourgues:2006nd,Hlozek:2016lzm,Komatsu:2008hk,Abdalla:2007ut,Tegmark:2005cy,Hu:2001bc,Abazajian:2016yjj,Moresco:2016nqq,Banerjee:2016suz,Archidiacono:2016lnv,Villaescusa-Navarro:2017mfx,Archidiacono:2016lnv,Hamann:2012fe,Yang:2017amu,Ruggeri:2017dda,Boyle:2017lzt,Zennaro:2017qnp,Doux:2017tsv,Wong:2011ip,Lattanzi:2017ubx,Slepian:2017dld}. 
Broadly, neutrinos transition from being a relativistic gas behaving as radiation in the early Universe to being a non-relativistic fluid that behaves like Cold Dark Matter (CDM). They decouple from the cosmic plasma while still relativistic, at which point they begin to free-stream. The horizon size corresponding to when neutrinos become non-relativistic thus sets a characteristic scale below which power appears suppressed.

Neutrinos therefore contribute to the total matter density, proportional to the sum of their masses, at late times, but with suppressed clustering at small scales. This is unlike CDM, which clusters strongly on all scales at low redshifts to aid in structure formation. These features can be measured through a relative amplitude measurement, or a  measurement of the change in shape of the matter power spectrum at small scales. Currently, growth measurements from \emph{Planck} combined with baryonic acoustic oscillations (BAO) from low-redshift surveys~\citep{Drinkwater:2009sd,Ross:2014qpa,Anderson:2013zyy} constrain the total neutrino mass to be  $\Sigma m_\nu \leq 0.21$ eV at 95\% confidence~\citep{Ade:2015xua}~\footnote{Note that the fiducial \emph{Planck} $\Lambda$CDM constraints assume neutrinos with a normal mass ordering and $\Sigma m_\nu \approx 60$ meV, dominated by the heaviest mass eigenstate.}. Here the \LCDM\ model is assumed, with a cosmological constant and no curvature. Constraints including different large scale structure observations have additionally been considered and can set still tighter bounds~\citep{Vagnozzi:2017ovm,Giusarma:2018jei,Gariazzo:2018pei,Palanque-Delabrouille:2015pga,Cuesta:2015iho,Giusarma:2016phn,DiValentino:2015sam,Couchot:2017pvz,Doux:2017tsv}.

The effect of massive neutrinos on CMB and LSS observables, in particular the suppression of clustering, can be mimicked by non-minimal extensions to $\Lambda$CDM cosmology, in particular through changes in dark energy and its dynamics, or changes in the spatial curvature, as shown in~\citep{Allison:2015qca}. Canonically in $\Lambda$CDM the ratio of dark energy pressure to its density is $w=-1$, but many models predict deviations from this constant value ~\cite[e.g.,][]{Doran:2006kp,Linder:2002et}, and $w\neq-1$ is consistent with current data~\citep{Ade:2015xua,Ade:2015rim,Abbott:2017wau,Alam:2016hwk}. 
We consider a description of possible dynamical dark energy using the standard Taylor expansion in the scale factor~\citep{Chevallier:2000qy,Linder:2002et}:
\begin{equation}
w(a)=w_0 + w_a(1-a),
\end{equation}
as well as non-zero spatial curvature. Due to the accelerated expansion, growth in dark matter structures and induced structure formation slows down in a dark energy-dominated universe. Varying the equation of state of the dark energy affects both the growth of structure and the expansion rate of the Universe, impacting both the amplitude of the power spectrum and the angular position of the baryon acoustic oscillations. Curvature also affects the same observables: varying the geometry modifies the angular diameter distances to a given redshift, and to hold fixed the primary CMB peak positions requires changing the matter density which affects the growth. 

\begin{figure}[h]
\centering
\includegraphics[width=0.4\textwidth]{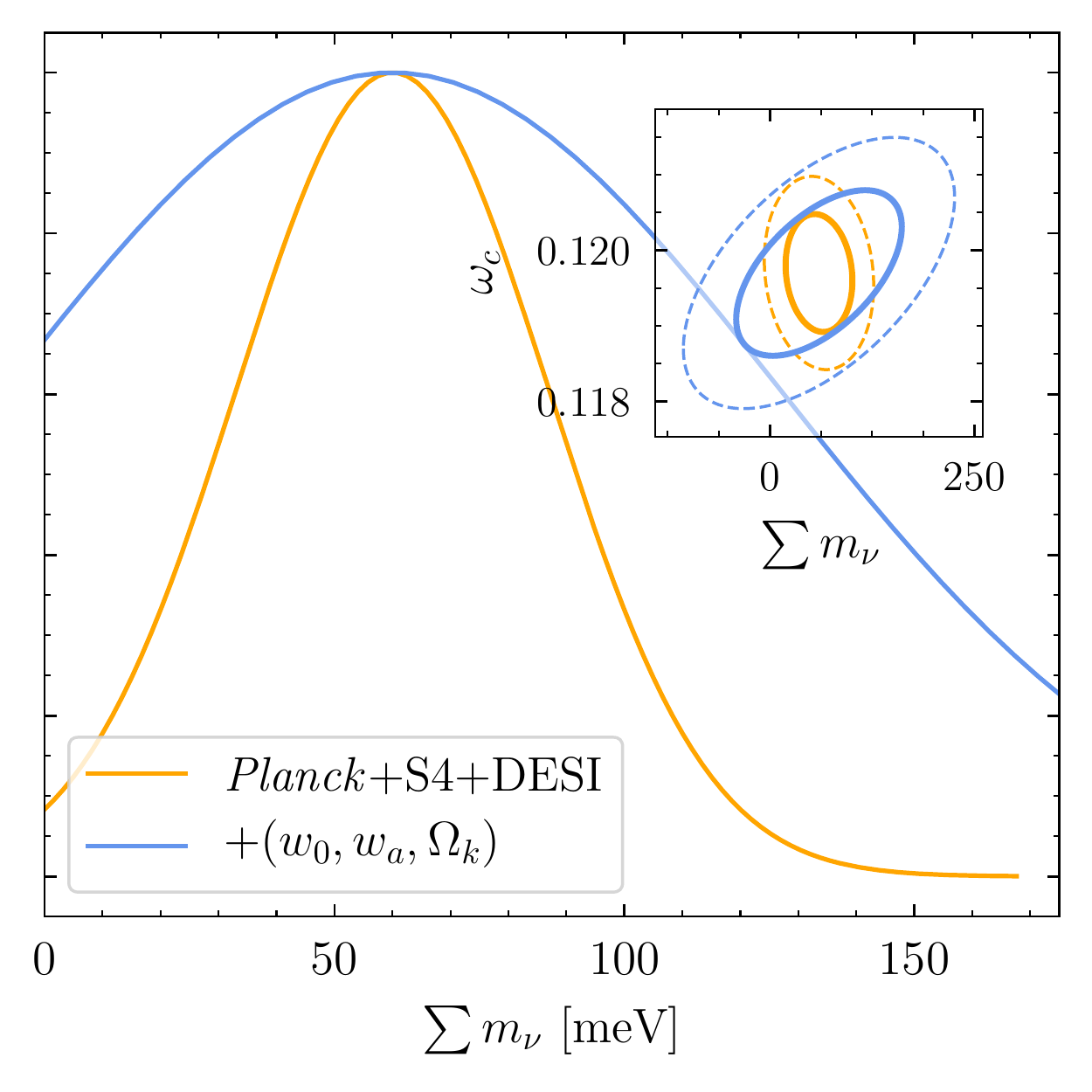}
\caption{Degradation in the error on the sum of neutrino masses with extensions to the canonical $\Lambda$CDM model, as in~\citet{Allison:2015qca}. Inset shows degeneracy with total CDM energy density when $w_0$, $w_a$ and $\Omega_k$ are opened up as free parameters. The forecast uncertainty $\sigma(\Sigma m_\nu)$ increases from 28 to 72 meV in this case.}
\label{fig:S4wok}
\end{figure}

If we had a measurement of the growth at only one single effective redshift, for example through CMB lensing, the degenerate effects of neutrino mass, dark energy equation of state, and curvature lead to a degradation in forecast sensitivity when these parameters are jointly varied. In Fig.~\ref{fig:S4wok} we show the forecast sensitivity on $\Sigma m_\nu$ with the same combination \emph{Planck}+CMB-S4+DESI considered in~~\citep{Allison:2015qca}, showing the effect of adding in dark energy and its linear evolution $w_0,w_a$ as well as curvature $\Omega_k$ as free parameters to the forecast neutrino mass. Here $\sigma(\Sigma m_\nu)$ is degraded to an uncertainty of $72$ meV compared to $\sigma(\Sigma m_\nu) = 28$ meV when these are not varied (see Tab.~\ref{tab:constraints} and~\citep{Allison:2015qca} for details). The addition of baryon acoustic oscillation measurements only partially breaks the degeneracies.

Accurate measurements of the late-time matter power spectrum at multiple epochs provide a path for disentangling the neutrino mass signal from non-minimal cosmological scenarios.
Gravitational lensing of background galaxies by foreground large-scale structure measures the growth of the projected matter distribution over several redshift bins, which can directly probe the suppression of structure at late times. Including the clustering of galaxies over cosmic time also adds both growth and distance information.

\section{Projected data and forecasting method}
\label{sec:datasets}

We use a standard Fisher forecasting method to project the expected constraints on cosmological parameters, using, as observables, angular power spectra of the primary CMB anisotropies, the CMB lensing convergence and projected LSST shear and galaxy clustering measurements. We use the \texttt{GoFish} code\footnote{Code available at \url{https://github.com/damonge/GoFish}. The results of the present analysis may be reproduced using the configuration files and \texttt{Jupyter} notebooks in the \texttt{LSST\_nu} folder.} originally described in~\citep{Alonso:2015sfa}, which has been refined and extended for this analysis and for internal forecasts for the CMB-S4 and LSST collaborations.

We forecast an extended set of ten free cosmological parameters:
\begin{equation}\nonumber
\{\Omega_bh^2, \Omega_ch^2, H_0, A_s, n_s, \tau, \Sigma m_{\nu}, w_0, w_a, \Omega_k\}~\footnote{We assume fiducial values of \{0.02222, 0.1197, 0.69, 2.1955, 0.9655, 0.06, 60, -1, 0, 0\} and step sizes of \{0.0001, 0.001, 0.01, 0.01, 0.005, 0.02, 10, 0.01, 0.05, 0.015\} for these parameters, and have verified the stability of the Fisher analysis to this choice.}.
\end{equation}
The first six are core \LCDM\ parameters: baryon density, cold dark matter density, Hubble constant, amplitude and slope of the primordial spectrum of metric fluctuations, and optical depth to reionization. The extension parameters are the neutrino mass sum $\Sigma m_{\nu}$, dark energy with equation of state $w(a)=w_0+(1-a)w_a$ and the curvature parameter $\Omega_k$. In places we will consider a limited subset of this model with fixed $\{w_0,w_a\}=\{-1,0\}$, or fixed $\Omega_k=0$. Models with more complicated equations of state, or with modifications to general relativity, in particular those that more closely mimic neutrinos or curvature, can be expected to introduce additional degeneracies~\citep{Lorenz:2017fgo, Dhawan:2017leu,Vagnozzi:2018jhn,Anand:2017ktp}.

Below we describe the specifications used for each experiment, and the associated observables and nuisance parameters. Where experiments are still in their design or proposal stages, we use nominal values proposed by the respective collaborations. Both collaborations have carried out their own forecasts, or are in the process of, producing their own science requirement studies, and our results should not be interpreted as official forecasts for either experiment.

\subsection{Cosmic Microwave Background}
\label{sec:CMB}

We use the proposed CMB Stage-4 (S4) experiment~\citep{Abazajian:2016yjj}, which will aim to observe half the sky in intensity and polarization, with anticipated white noise levels of 1 $\mu$K-arcmin in intensity. We assume that 40\% of the sky will be useable for analysis, and that the CMB lensing field will be reconstructed in the angular range $\ell\in[50,5000]$. We also consider experiments with higher noise levels increasing up to 10 $\mu$K-arcmin, which includes the level anticipated for the Simons Observatory\footnote{\url{https://simonsobservatory.org/}}. Since CMB-S4 is a ground-based experiment, characterized by a large, non-white noise contribution on large scales, we use a minimum multipole $\ell=50$, and add large-scale temperature and polarization information from \emph{Planck}~\citep{Ade:2015xua} below that scale. The noise levels have been calibrated to reproduce the errors reported in the 2015-16 \planck\ papers, with the large-scale polarization noise tuned to reproduce an optical depth uncertainty of $\sigma(\tau)\approx 0.01$~\citep{Aghanim:2016yuo}. In cases where we do not use CMB-S4 in forecasts, we extend the \emph{Planck} multipole range up to $\ell=3000$. 
We also consider the impact of adding forecast constraints from proposed space-based experiments (\emph{e.g.} LiteBIRD~\citep{Matsumura:2013aja}) that better measure the optical depth from large-scale polarized $E$-modes. The specifications are summarized in Tab.~\ref{tab:cmbprim}. The polarization noise level of 4 $\mu$K-arcmin is conservatively higher than anticipated from LiteBIRD, but still able to give a cosmic-variance limited measurement of $\tau$.

\begin{table}[h]
\begin{center}
\begin{tabular}{cccccc}
\hline \hline
Experiment & $\ell$-range & $f_{\rm sky}$ & Beam & $\Delta T$ &  $\Delta P$ \\ 
&  & & [arcmin] & [$\mu$K-arcmin] &  [$\mu$K-arcmin] \\ 
\hline  
\emph{Planck} & [2,50] & $0.4$ & 10 & 31.1 & 150.0  \\
S4 & [50,5000] &  $0.4$ &  3  & 1.0 & 1.4    \\
CV  & [2,50] &  $0.7$ &  30 & 2.8   & 4.0     \\
\hline
\end{tabular}
\end{center}
\caption{Specifications of the CMB experiments used in our forecasts. When S4 is excluded, the range of \emph{Planck} is extended to $\ell=3000$ with $\Delta T = 31.1\,\mu$K-arcmin and   $\Delta P = 56.9\,\mu$K-arcmin at higher multipoles, calibrated to reproduce the constraints in~\citep{Ade:2015xua}. The `CV' experiment has non-zero noise but is cosmic variance limited in polarized $E$-modes on large-scales.}
\label{tab:cmbprim}
\end{table}

\subsection{Large-scale structure from LSST} 
The Large Synoptic Survey Telescope (LSST)~\citep{Abell:2009aa} is a Stage-IV photometric survey that will map out the galaxy distribution on half the sky down to magnitude $r\sim27$. LSST will pursue 5 main cosmological observables: weak lensing, galaxy clustering, clusters of galaxies, supernovae and strong lensing. Of these, we focus here on the first two. We assume that the baseline data vector for LSST will be in the form of a ``$3\times2$-pt'' analysis~\citep{Krause:2017ekm}, based on the combination of weak lensing and galaxy clustering, both in auto-correlation and cross-correlation. The specifications of the lensing and clustering samples used here follow closely those assumed in~\citep{Alonso:2015sfa}, which we describe briefly below.

\paragraph{Galaxy clustering.} We divide the galaxy clustering sample into two populations of ``red'' and ``blue'' galaxies. The red sample is characterized by better photometric redshift (photo-$z$) uncertainties (see below), lower number density, a higher galaxy bias and shallower redshift coverage. The blue sample, on the other hand, has a much higher number density and reaches to higher redshifts, at the cost of increased photo-$z$ uncertainties. The redshift distribution of these samples is based on the measurements of the luminosity functions presented by~\citep{Faber:2005fp} (red sample) and~\citep{Gabasch:2005bb} (all galaxies), and assuming a magnitude limit $i\sim25.3$, corresponding to the so-called LSST ``gold'' sample~\citep{Abell:2009aa}. The $k$-corrections needed to transform the luminosity functions into redshift distributions were computed using {\tt kcorrect}~\citep{Blanton:2006kt}. We model the photo-$z$ distribution of both samples as a Gaussian with a redshift-dependent standard deviation $\sigma_z=\sigma_0\,(1+z)$, with $\sigma_0=0.02$ and $0.05$ for the red and blue samples respectively. Finally, we split the full redshift range covered by each sample into bins distributed such that the width of each bin in photo-$z$ space is three times the value of $\sigma_z$ at the bin centre. This results in 15 bins for the red sample and 9 bins for the blue sample. For more details on the clustering specification, see~\citep{Alonso:2015uua}. Our fiducial constraints will only include the blue sample, but we will also explore the impact of adding the red sample in Section \ref{sec:systematics}.

\paragraph{Cosmic shear.} The distribution of inhomogeneities in the cosmic density field produces perturbations on the trajectories of photons emitted by distant sources, an effect known as gravitational lensing~\citep{Bartelmann:1999yn}. This effect can be traced by the correlated distortions it produces in the observed shapes of galaxies, labelled ``cosmic shear'' or ``weak lensing'', and is the same effect probed by the lensing of the CMB. As a direct probe of the density fluctuations, cosmic shear is a tremendously useful cosmological probe, particularly in combination with galaxy clustering. The LSST weak lensing sample is modeled here using the ``fiducial'' redshift distribution estimated by~\citep{Chang:2013xja}. As in the case of galaxy clustering we assume a tomographic analysis in which the sample is split into redshift bins. For this, we model the photo-$z$ uncertainty of the lensing sample as Gaussian with $\sigma_z=0.05\,(1+z)$, and define 9 top-hat bins in photo-$z$ space with a width corresponding to $3\times\sigma_z$.

\paragraph{Angular power spectra.}
The data vector considered for LSST is the collection of maps of the galaxy overdensity or the cosmic shear field associated with the samples described above. Let $a^\alpha_{\ell m}$ be the harmonic coefficients of the $\alpha$-th map, corresponding for instance to one of the galaxy clustering redshift bins, and let us collect all maps for a given multipole order $(\ell,m)$ into a vector ${\bf a}_{\ell m}$. The power spectrum ${\sf C}_{\ell}$ is defined as the covariance of this vector $\langle {\bf a}_{\ell m}{\bf a}^\dag_{\ell' m'}\rangle\equiv{\sf C}_\ell\delta_{\ell\ell'}\delta_{mm'}$. For both galaxy clustering and cosmic shear, the cross-correlation between two maps $C^{\alpha\beta}_\ell$ can be directly related to the 3D matter power spectrum as
\begin{equation}
  C^{\alpha\beta}_\ell=\frac{2}{\pi}\int_0^\infty dk\,k^2\,\Delta^\alpha_\ell(k)\Delta^\beta_\ell(k),
\end{equation}
where $\Delta^\alpha_\ell(k)$ is a transfer function associated to the $\alpha$-th map.

For the galaxy overdensity in the $\alpha$-th redshift bin, this is given by~\citep{Alonso:2015uua,Alonso:2015sfa,DiDio:2013bqa}
\begin{align}\nonumber
  &\Delta^{{\rm GC},\alpha}_\ell(k)\equiv\int dz\,\phi^\alpha(z)\,\Psi_\ell(k,z)\,\sqrt{P(k,z)},\\
  &\Psi_\ell(k,z)\equiv b^\alpha(z)j_\ell(k\chi(z))-f(z)j_\ell^{''}(k\chi(z)),
\end{align}
where $\phi^\alpha(z)$ is the selection function of the $\alpha$-th bin, $b^\alpha(z)$ is the linear galaxy bias, $f\equiv d\log\delta/d\log a$ is the growth rate, $j_\ell(x)$ is the spherical bessel function of order $\ell$, $\chi(z)$ is the radial comoving distance to redshift $z$ and $P(k,z)$ is the matter power spectrum.

For cosmic shear\footnote{Note that the cosmic shear is a spin-2 field, which can be decomposed into $E$ and $B$ modes. We only include $E$ modes in our analysis, which is the only non-zero contribution for the cosmological models explored here.}, the transfer function for the $\alpha$-th redshift bin is given by
\begin{align}\nonumber
 &\Delta^{{\rm CS},\alpha}_\ell(k)\equiv\sqrt{\frac{(\ell+2)!}{(\ell-2)!}}\int d\chi\,W^\alpha(\chi)\frac{j_\ell(k\chi)}{k^2a(\chi)}\sqrt{P(k,z)},\\
 &W^\alpha(\chi)\equiv\frac{3\Omega_{M,0}H_0^2}{2}\int_{z(\chi)}^\infty dz'\phi^\alpha(z')\frac{\chi(z')-\chi}{\chi\chi(z')}
\end{align}
We also include all the angular cross-spectra between CMB lensing, cosmic shear, and galaxy clustering.

The noise power spectra for clustering and shear are associated to the discrete sampling of the underlying density field and the intrinsic scatter of galaxy shapes. They are zero for disjoint redshift bins, and their contribution to the auto-correlation is given by $N^{i,{\rm clust}}_\ell=1/\bar{n}_i$ and $N^{i,{\rm shear}}_\ell=\sigma_\gamma^2/\bar{n}_i$, where $\bar{n}_i$ is the projected number density of sources (in units of srad$^{-1}$) and $\sigma_\gamma=0.28$ is the intrinsic shape noise per ellipticity component~\citep{Abell:2009aa}.

\begin{figure*}[htbp]
  \centering
  \includegraphics[width=0.49\textwidth]{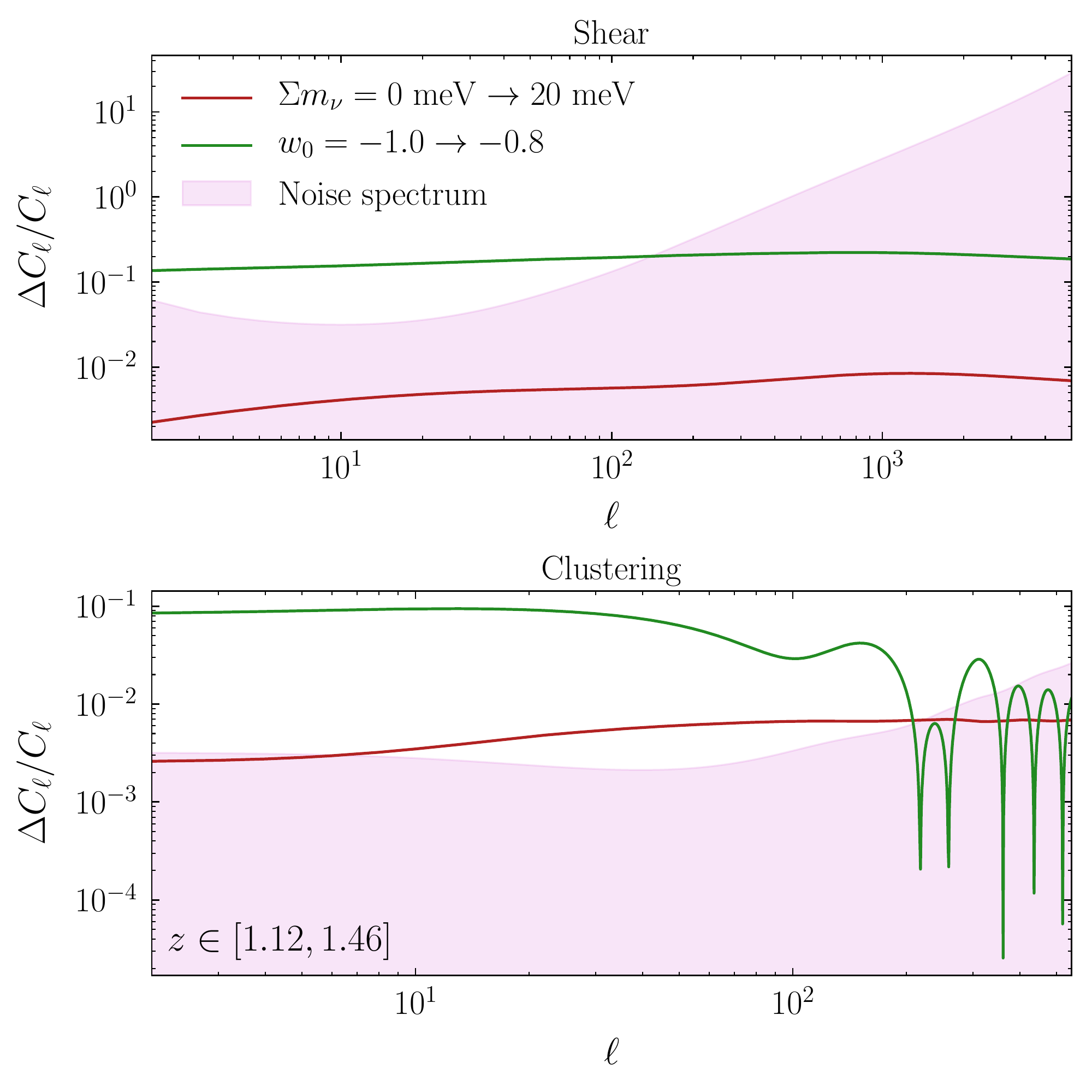}
  \includegraphics[width=0.48\textwidth]{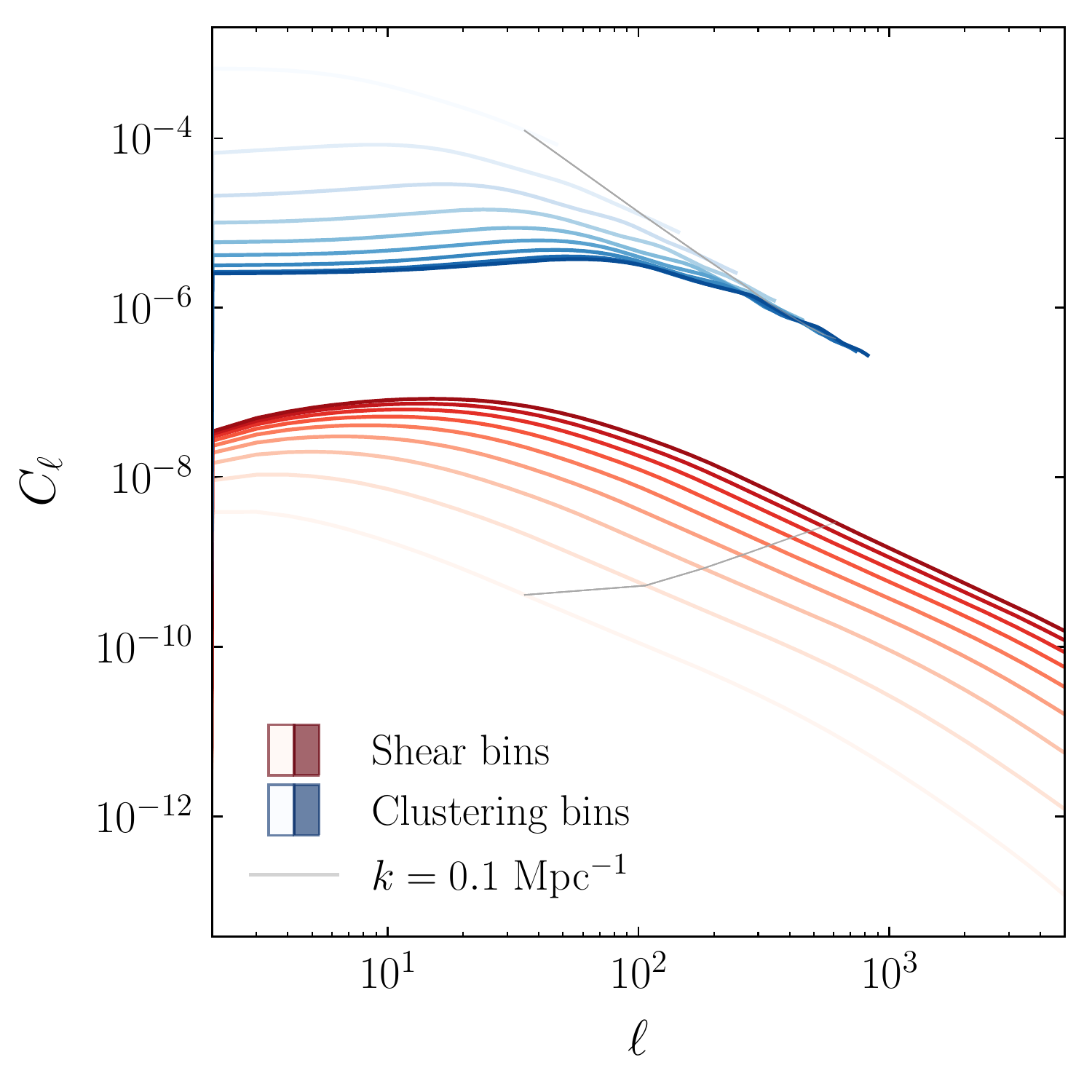}
  \caption{(Left) The relative change in the power spectrum for the case of cosmic shear (top) and galaxy clustering (bottom) when the total neutrino mass and $w_0$ are varied, corresponding to Bin 5 of our setup ($z \in [1.12, 1.46]$). Also shown for comparison is the LSST noise spectrum considered in this redshift range. (Right) The auto-power spectra for the fiducial cosmological parameters, shown for all the shear (red) and clustering (blue) redshift bins considered, including scale cuts. Darker colours correspond to higher redshift bins. For reference, the grey line correspond to a scale of $k = 0.1$ Mpc$^{-1}$.}
  \label{fig:Cls}
\end{figure*}

\paragraph{Systematic effects.}

Both galaxy clustering and cosmic shear suffer from a number of sources of systematic uncertainties that introduce extra nuisance parameters that must be marginalized over. We review these here:
\begin{itemize}
\item{\emph{Galaxy bias:}} The relation between the galaxy and matter power spectra is expected to be well-approximated by a linear, scale-independent, factor $b(z)$ on large scales. Our forecasts therefore marginalize over the value of this quantity defined, for each galaxy sample, at a discrete set of nodes in redshift (with the full $b(z)$ function reconstructed by interpolating between these nodes). See~\citep{Alonso:2015uua} for further details. As will be discussed i in the last point below, we use redshift-dependent scale cuts to mitigate against the effects of scale-dependent bias terms~\citep{Chiang:2017vuk,Giusarma:2018jei,Raccanelli:2017kht}.
\item{\emph{Baryonic effects:}} We use the prescription in~\citep{Schneider:2015wta} to account for the effects of baryons in the angular power spectra using their three-parameter correction model. The parameters and their fiducial values are the mass dependence of the halo gas fraction (log$_{10}(M_{bcm}/M_\odot h^{-1})=14.08$), the corresponding ejection radius as a fraction of the virial radius ($\eta_b=0.5$) and the scale of the stellar component ($k_s=55\,h\,{\rm Mpc}^{-1}$), which we additionally marginalize over. 
\item{\emph{Multiplicative bias:}} Estimating the shear from galaxy shapes may lead to redshift-dependent multiplicative biases~\citep{Schaan:2016ois,Huterer:2005ez,Massey:2012cz}. The shear multiplicative bias is degenerate with the amplitude of the signal and its time evolution can hide the true evolution of the growth of structure, which probes dark energy as well as massive neutrinos. We apply an overall multiplicative factor $(1+m_\alpha)$ in each shear redshift bin $\alpha$, marginalizing over each $m_\alpha$. We use fiducial values of zero and step sizes of 0.005 for $m_\alpha$ in the Fisher analysis.
\item{\emph{Intrinsic alignments:}} Intrinsic alignments (IA) of galaxies associated to structure formation can contaminate the cosmic shear signal by up to 1-10\%~\citep{Krause:2015jqa,Schaan:2016ois,Troxel:2014dba,Kiessling:2015sma,Kirk:2015nma,Joachimi:2015mma}. Evidence suggests that IA are caused to a large extent by the alignment of galaxies with the direction of tidal forces in the cosmic web. We model this effect according to the so-called non-linear alignment model~\citep{Hirata:2004gc}, and marginalize over 4 values describing the redshift evolution of the overall amplitude of the IA contribution.
\item{\emph{Photo-$z$ uncertainties:}} Inaccuracies in the characterization of the photo-$z$ distribution of individual sources lead to uncertainties in the overall galaxy redshift distribution in each redshift bin, needed to estimate theoretical predictions for the expected power spectrum. We therefore marginalize over two additional parameters in each clustering redshift bin -- an overall bias in the determination of photo-$z$s as well as over the scatter of redshifts~\citep{Schaan:2016ois}. We impose a prior of 0.005 (corresponding to the width of a Gaussian prior centered on the fiducial value of zero) on the photo-$z$ bias following~\citep{Abbott:2017wau} and with the expectation that LSST will be able to achieve better sensitivity.
\item{\emph{Scale cuts:}} Theoretical uncertainties in the effect of baryons on the matter power spectrum, and on the details of the galaxy-matter connection prevent us from using the smallest scales of the shear and galaxy power spectrum respectively. For this reason we must drop all multipoles beyond a given $\ell_{\rm max}$. For cosmic shear, where the main effect is that of baryons, we choose a fiducial cut $\ell_{\rm max}=5000$, independent of redshift. This corresponds to a comoving scale $k_{\rm max}\sim1.6\,h\,{\rm Mpc}^{-1}$ at $z\sim1.5$. Since galaxy clustering is affected by complicated non-linear, non-local and scale-dependent bias terms~\citep{Chiang:2017vuk,Giusarma:2018jei,Raccanelli:2017kht}, we use a more conservative, redshift-dependent cut. In this case, at the median redshift of each bin $\bar{z}$, we compute a minimum comoving scale $k_{\rm max}(\bar{z})$ defined by requiring that matter perturbations up to that wavenumber have a standard deviation $\sigma(<k_{\rm max})=0.75$. We then translate $k_{\rm max}(\bar{z})$ into an angular scale $\ell_{\rm max}(\bar{z})=\chi(\bar{z})\,k_{\rm max}(\bar{z})$. We do not impose additional cuts at large scales. The resulting scale cuts can be seen in the left panel of Fig. \ref{fig:Cls}, which shows the auto-power spectra of all redshift bins for clustering and lensing considered here.
\end{itemize}

\begin{figure*}[htbp]
  \centering
  \includegraphics[width=0.3\textwidth]{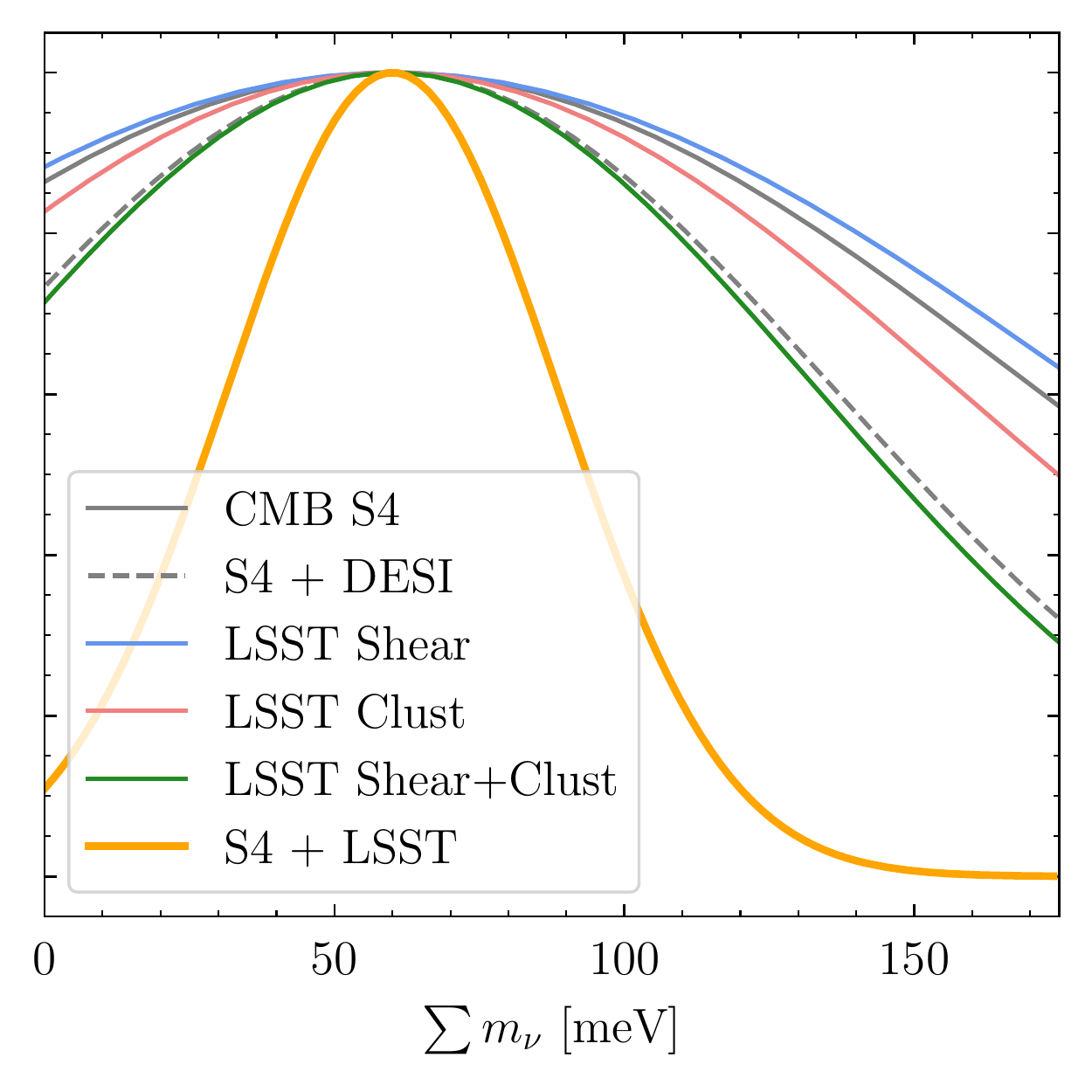}
  \includegraphics[width=0.3\textwidth]{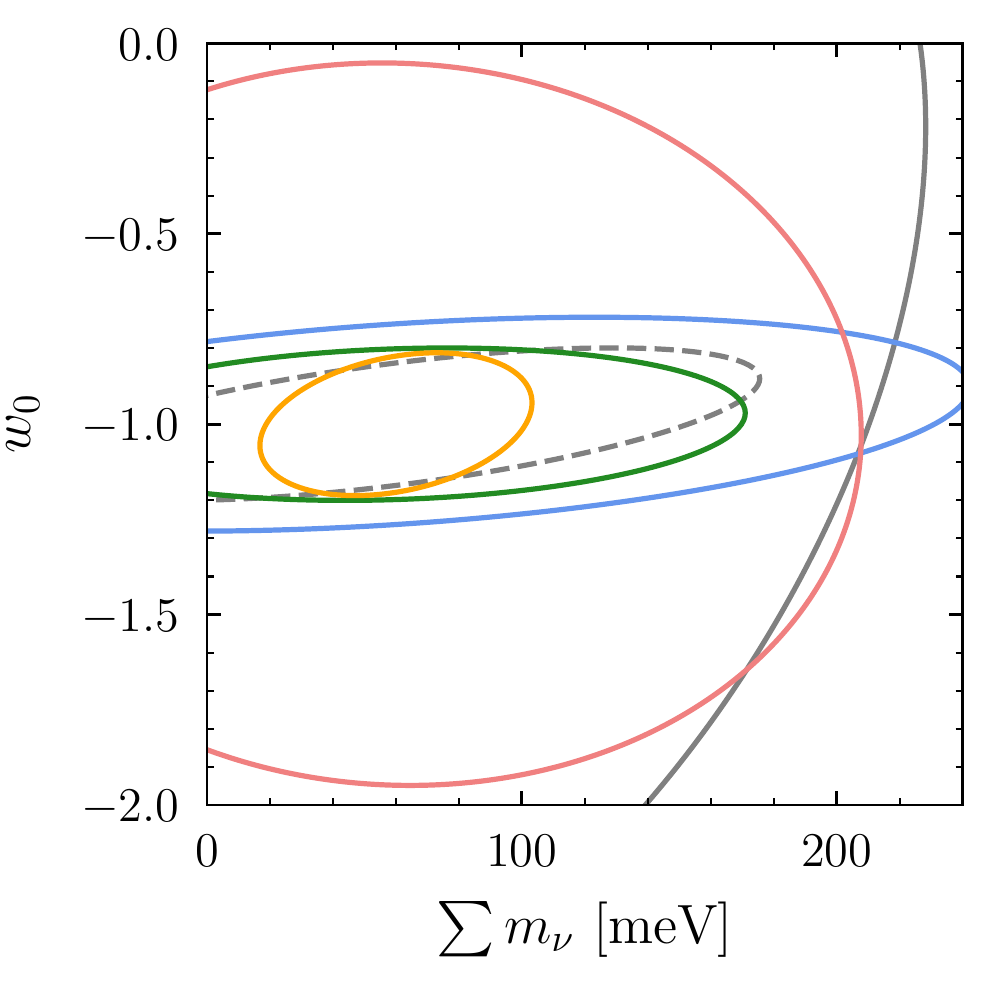}
  \includegraphics[width=0.3\textwidth]{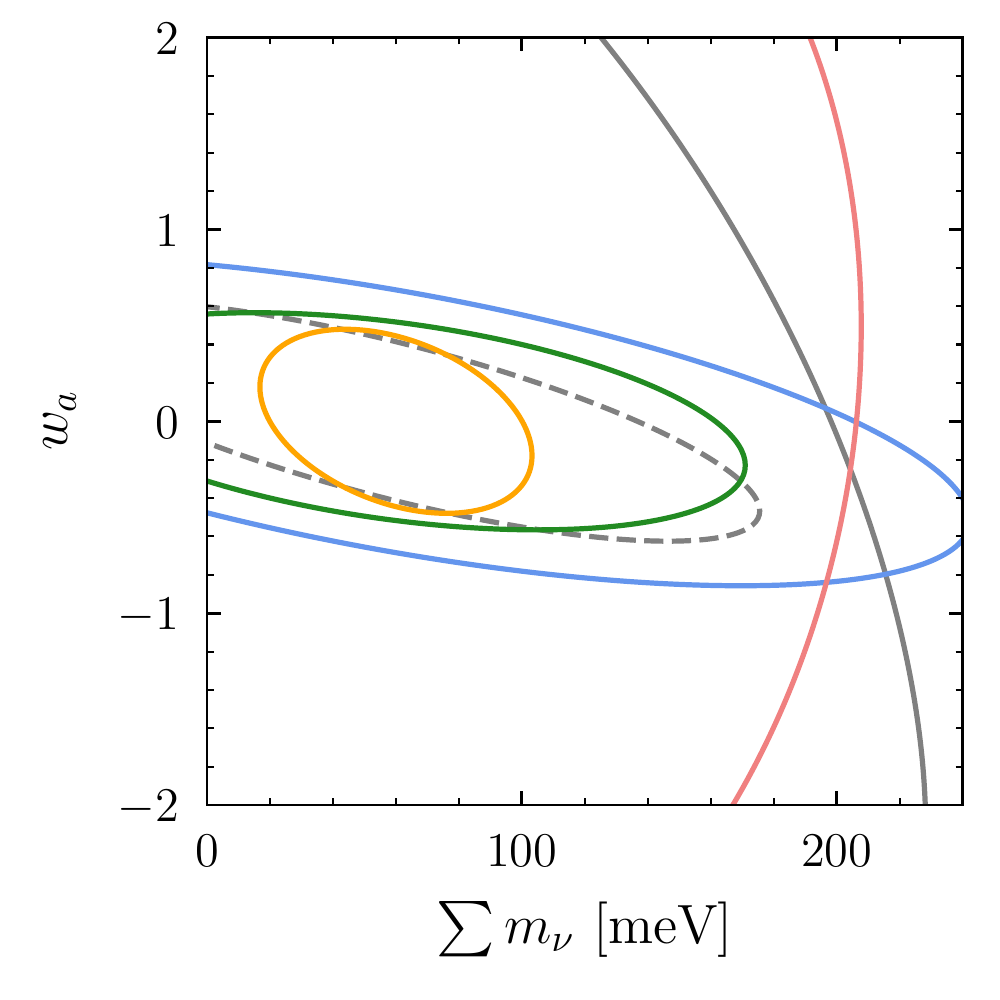}

  \caption{{\sl Left:} Forecast error on $\Sigma m_\nu$ achievable with CMB-S4 (grey), LSST shear (blue), LSST clustering (red), LSST clustering and shear (green) and all together (orange), combined with \emph{Planck} primary CMB data as described in Sec.~\ref{sec:CMB}, in the presence of an uncertain dark energy equation of state. {\sl Center, right:} Forecast error on $w_0$ and $w_a$ with different combinations of probes, revealing the degeneracies with $\Sigma m_\nu$ in each case. The corresponding forecast values are given in Tab.~\ref{tab:constraints}.} \label{fig:comp}
\end{figure*}

\subsection{Baryon Acoustic Oscillation from DESI}
Besides the CMB and large-scale structure datasets described above, we will also include in some cases the expected constraints from the measurements of the redshift-distance relation made with the future Dark Energy Spectroscopic Instrument (DESI) due to begin in 2018~\citep{Levi:2013gra}. We use the forecast uncertainties on $d_A(z)$ and $H(z)$ provided in~\citep{Aghamousa:2016zmz}, in 13 redshift bins between $z=0.65$ and 1.85 and 5 redshift bins between $z=0.05$ and 0.45 with bin width $\Delta z=0.1$, expected to be achievable by DESI and the DESI Bright Galaxy Survey respectively covering $14,000~\mathrm{deg}^2$. 

\section{Results}
\label{sec:results}

In this section, we study the sensitivity of LSST and a future CMB-S4 experiment to the sum of neutrino masses, the dark energy equation of state and cosmological curvature. We consider the following combinations:

\begin{itemize}
\item Setup 1: \emph{Planck} + S4 (+ DESI BAO, as in~\citep{Allison:2015qca})
\item Setup 2: \emph{Planck} + LSST-shear
\item Setup 3: \emph{Planck} + LSST-clustering
\item Setup 4: \emph{Planck} + LSST-clustering + LSST-shear
\item Setup 5: \emph{Planck} + S4 + LSST-shear + LSST-clustering
\end{itemize}
We also consider in this section the impact of including a cosmic variance (CV)-limited measurement of the optical depth to reionization $\tau$, and describe the effect of including BAO measurements from DESI as an additional tracer of late-time clustering.

\subsection{Forecasts with LSST and CMB-S4}
\label{subsec:mnucomplementarity}

\begin{figure}[h]
\centering
\includegraphics[width=0.48\textwidth]{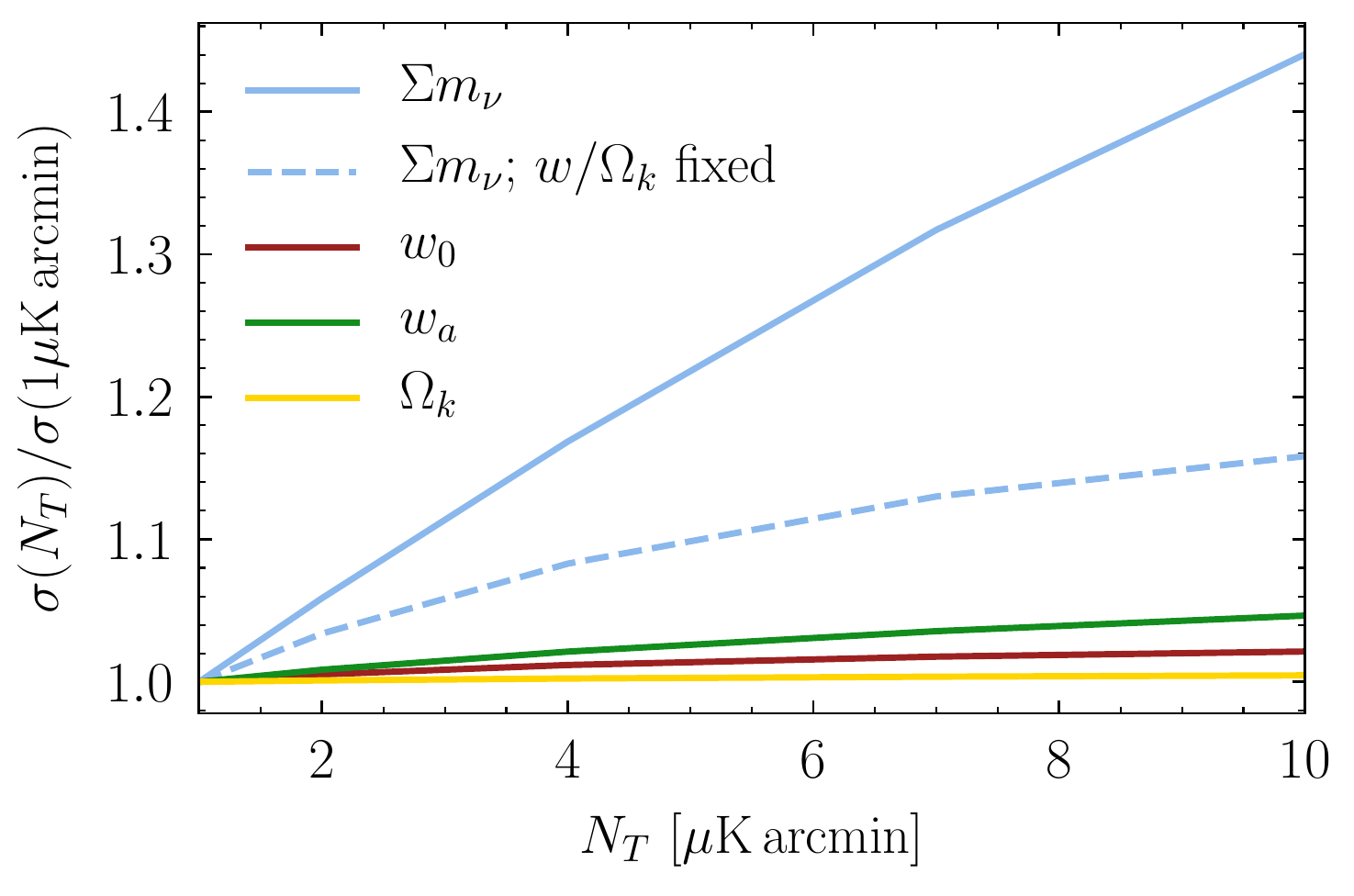}
\caption{Achievable constraints on $\Sigma m_\nu$ (blue), $w_0$ (burgundy), $w_a$ (green) and $\Omega_k$ (yellow) as a function of the CMB noise level in intensity $N_T$. Forecasts are shown as a ratio to the constraints achievable for a $1\ukam$ experiment. Although $w_0$, $w_a$ and $\Omega_k$ do not degrade significantly with $N_T$, the uncertainty on the sum of neutrino masses could improve by $\sim40\%$ from a Stage-3 experiment ($\sim10\ukam$) to S4. Also shown (dotted blue) are the achievable constraints on $\Sigma m_\nu$ when $w_0$, $w_a$ and $\Omega_k$ are fixed to their fiducial \LCDM~values. The relative degradation with increasing CMB noise level is much more modest in this case.}
\label{fig:mnu_st}
\end{figure}

Figure~\ref{fig:comp} shows forecast constraints for the error on $\Sigma m_\nu$, the dark energy parameters $w_0$, $w_a$ and the cosmological curvature $\Omega_k$ obtainable with shear and clustering measurements from LSST and CMB-S4. These results are shown in Tab.~\ref{tab:constraints}. In all these cases \emph{Planck} is included as described in Sec \ref{sec:datasets}. Individually, CMB-S4, LSST clustering and LSST shear can achieve forecast constraints of $\sigma(\Sigma m_\nu)=$ 111, 91 and 120 meV respectively, strongly degraded with respect to the case where the flat \LCDM\ fixed-$w$ model is assumed ($\sigma(\Sigma m_\nu)=$ 73, 69 and 41 meV respectively). In combination, however, the three probes are able to achieve an error of $\sigma(\Sigma m_\nu)=28$ meV. This would be an almost 4$\sigma$ measurement of the minimal mass in the inverted hierarchy, and $\approx2\sigma$ for the normal hierarchy. By combining these datasets, the degradation is only $\sim20\%$ with respect to the fixed-\LCDM\ case. 

It is worth pointing out that, while a free equation of state $w\neq-1$ represents a more complex extension of the standard $\Lambda$CDM model, in which the accelerated expansion is driven by something other than a simple cosmological constant, the spatial curvature $\Omega_k$ is a core parameter needed to fully describe the Friedmann-Robertson-Walker metric. It is therefore interesting to investigate the possible degeneracy on $\Sigma m_\nu$ from freeing $\Omega_k$ while fixing $(w_0,w_a)$. This case is shown in the last row of Tab.~\ref{tab:constraints}: $\Omega_k$ is significantly less degenerate with $\Sigma m_\nu$, and therefore the uncertainty on the latter parameter remains unchanged after freeing up the former.

The improvement on $\sigma(\Sigma m_\nu)$ from the combination of CMB and LSS probes is a result of the high sensitivity of future CMB data. Experiments including Advanced ACTPol and the Simons Observatory will measure the CMB lensing over large sky areas to higher noise levels than S4, so we explore how the forecast constraint on neutrino masses depends on the CMB noise level, keeping the sky area fixed to 40\%. Figure \ref{fig:mnu_st} shows the relative degradation in the 1$\sigma$ uncertainty on $\Sigma m_\nu$, $\Omega_k$, $w_0$ and $w_a$ as a function of the CMB noise level in temperature $N_T$, with respect to the fiducial case $N_T=1\,\ukam$. We find that $\sigma(m_\nu)$ could be improve by $\sim40\%$ with respect to the constraints achievable with a Stage-3 experiment such as Advanced ACTPol~\citep{DeBernardis:2016uxo} ($N_T\approx10\ukam$), but the curvature and dark energy parameters would benefit little ($\lesssim5\%$ or less) from the higher sensitivity. Interestingly, we also find that the improvement in uncertainty on $\Sigma m_\nu$ with $N_T$ is much more modest when $\{w_0,\,w_a,\,\Omega_k\}$ are fixed, consistent with~\citet{Allison:2015qca}.

\begin{table}[h]
\begin{center}
\begin{tabular}{c|c|cccc}
\hline \hline
Setup & $\sigma(\Sigma m_\nu)$ & $\sigma(\Sigma m_\nu)$ & $\sigma(\Omega_k)$ &$\sigma(w_0)$ & $\sigma(w_a)$  \\ 
&  [meV]& [meV]& [$\times 10^{-3}$] &  \\ 
\hline  
S4 & 73 & 111 & 0.79 & 1.14 & 2.46 \\
( + DESI BAO) & 29 & 76 & 0.48 & 0.13 & 0.41 \\
\hline
LSST-clustering & 69 & 91 & 3.33 & 0.42 & 1.22 \\
LSST-shear & 41 & 120 & 2.99 & 0.19 & 0.57 \\
LSST-shear+clust & 32 & 72 & 2.06 & 0.11 & 0.33 \\
\hline
S4+LSST & 23 & 28 & 0.49 & 0.10 & 0.26 \\
& - & 24 & 0.49 & - & - \\
\hline
\end{tabular}
\end{center}
\caption{Forecast constraints on $\Sigma m_\nu$ from various combinations of probes combined with \emph{Planck} primary CMB data as described in Sec.~\ref{sec:CMB}. The first column assumes the \LCDM\ model. The second allows for degeneracies with the spatial curvature and a two-parameter dark energy equation of state. The minimal mass sum in a normal hierarchy is $\Sigma m_\nu\approx$ 60~meV, and $\Sigma m_\nu\approx$ 100~meV in an inverted hierarchy.}
\label{tab:constraints}
\end{table}

\subsection{Impact of extended datasets}
\label{subsec:cvlim_bao}

\noindent {\bf Optimal measurement of the optical depth}\newline
In Table~\ref{tab:constraints_cv} we show the forecast constraint on $\Sigma m_\nu$, the dark energy parameters $w_0$, $w_a$ and the cosmological curvature $\Omega_k$ obtainable when adding a cosmic-variance limited measurement of the optical depth to reionization with $\sigma(\tau)\approx0.002$.  In this case, LSST and CMB-S4 together are projected to achieve an error of $\sigma(\Sigma m_\nu)=21$~meV, enough for a measurement of the minimal neutrino mass at $3\sigma$ significance even within this broader cosmological model. These results show that an improved measurement of $\tau$ is vital to break the degeneracy with the amplitude of scalar perturbations, not only for CMB-based measurements as found in~\citet{Allison:2015qca}, but also for large-scale structure surveys aiming to constrain neutrino mass. We also note that, in the absence of S4, LSST alone would benefit less from a better measurement of $\tau$, projecting only a minimal improvement on $\sigma(\Sigma m_\nu)$. Finally, we find that improving the optical depth measurement has little impact on the $w_0$,  $w_a$ and $\Omega_k$ forecast constraints.
\\

\begin{table}[h]
\begin{center}
\begin{tabular}{c|c|cccc}
\hline \hline
Setup & $\sigma(\Sigma m_\nu)$ & $\sigma(\Sigma m_\nu)$ & $\sigma(\Omega_k)$ &$\sigma(w_0)$ & $\sigma(w_a)$  \\ 
(+CV-$\tau$) &  [meV]& [meV]& [$\times 10^{-3}$] &  \\ 
\hline  
LSST-clustering & 69 & 91 & 3.3 & 0.42 & 1.20 \\
LSST-shear & 31 & 117 & 2.82 & 0.18 & 0.55 \\
LSST-shear+clust & 24 & 72 & 1.99 & 0.11 & 0.31 \\
\hline  
S4+LSST & 14 & 21 & 0.49 & 0.10 & 0.26 \\
 & - & 15 & 0.49 & - & - \\

\hline
\end{tabular}
\end{center}
\caption{Forecast constraints on $\Sigma m_\nu$ as in Tab.~\ref{tab:constraints} but including a cosmic variance-limited $\tau$ measurement matching LiteBIRD sensitivity.}
\label{tab:constraints_cv}
\end{table}

\noindent {\bf Additional BAO measurements}\newline
Primordial oscillations in the baryon-photon fluid imprint characteristic geometric information in the distribution of galaxies, known as Baryon Acoustic Oscillations (BAO). Massive neutrinos are sensitive to the BAO scale through the angular diameter distance $d_A(z)$ and expansion rate $H(z)$. While galaxy clustering as measured by LSST will implicitly measure the BAO scale, uncertainties and systematics associated with the measured photometric redshifts will inevitably degrade any BAO-related constraints. It is therefore interesting to explore whether including measurements of the angular diameter distance and expansion rate redshift evolution from spectroscopic surveys would lead to a significant improvement in the final constraints. Table~\ref{tab:constraints_bao} shows forecast uncertainties on $\Sigma m_\nu$, the dark energy parameters $w_0$, $w_a$ and the cosmological curvature $\Omega_k$ obtainable with an additional BAO measurement from DESI. By comparison with Tab.~\ref{tab:constraints}, we observe that, although the additional constraining power of DESI's BAO measurements could significantly help some individual probes (e.g. LSST shear or clustering), the final combined constraints on $\Sigma m_\nu$ are only marginally improved by $\lesssim10\%$ by the additional information. Finally, constraints on $\Omega_k$  are improved by $\sim10\%$ and those on $w_0,w_a$ by $\sim20\%$ in this case. 

\begin{table}[h]
\begin{center}
\begin{tabular}{c|c|cccc}
\hline \hline
Setup & $\sigma(\Sigma m_\nu)$ & $\sigma(\Sigma m_\nu)$ & $\sigma(\Omega_k)$ &$\sigma(w_0)$ & $\sigma(w_a)$  \\ 
(+DESI BAO) &  [meV]& [meV]& [$\times 10^{-3}$] &  \\ 
\hline  
LSST-clustering & 36 & 89 & 1.72 & 0.14 & 0.44 \\
LSST-shear & 32 & 100 & 1.45 & 0.12 & 0.38 \\
LSST-shear+clust & 27 & 67 & 1.24 & 0.09 & 0.27 \\
\hline
S4+LSST & 20 & 28 & 0.45 & 0.08 & 0.20 \\
 & - & 22 & 0.44 & - & - \\
\hline
\end{tabular}
\end{center}
\caption{Forecast constraints on $\Sigma m_\nu$ as in Tab.~\ref{tab:constraints} but including projected DESI BAO measurements.}
\label{tab:constraints_bao}
\end{table}

\subsection{Impact of systematics}
\label{sec:systematics}
\begin{figure}[h]
\centering
\includegraphics[width=0.5\textwidth]{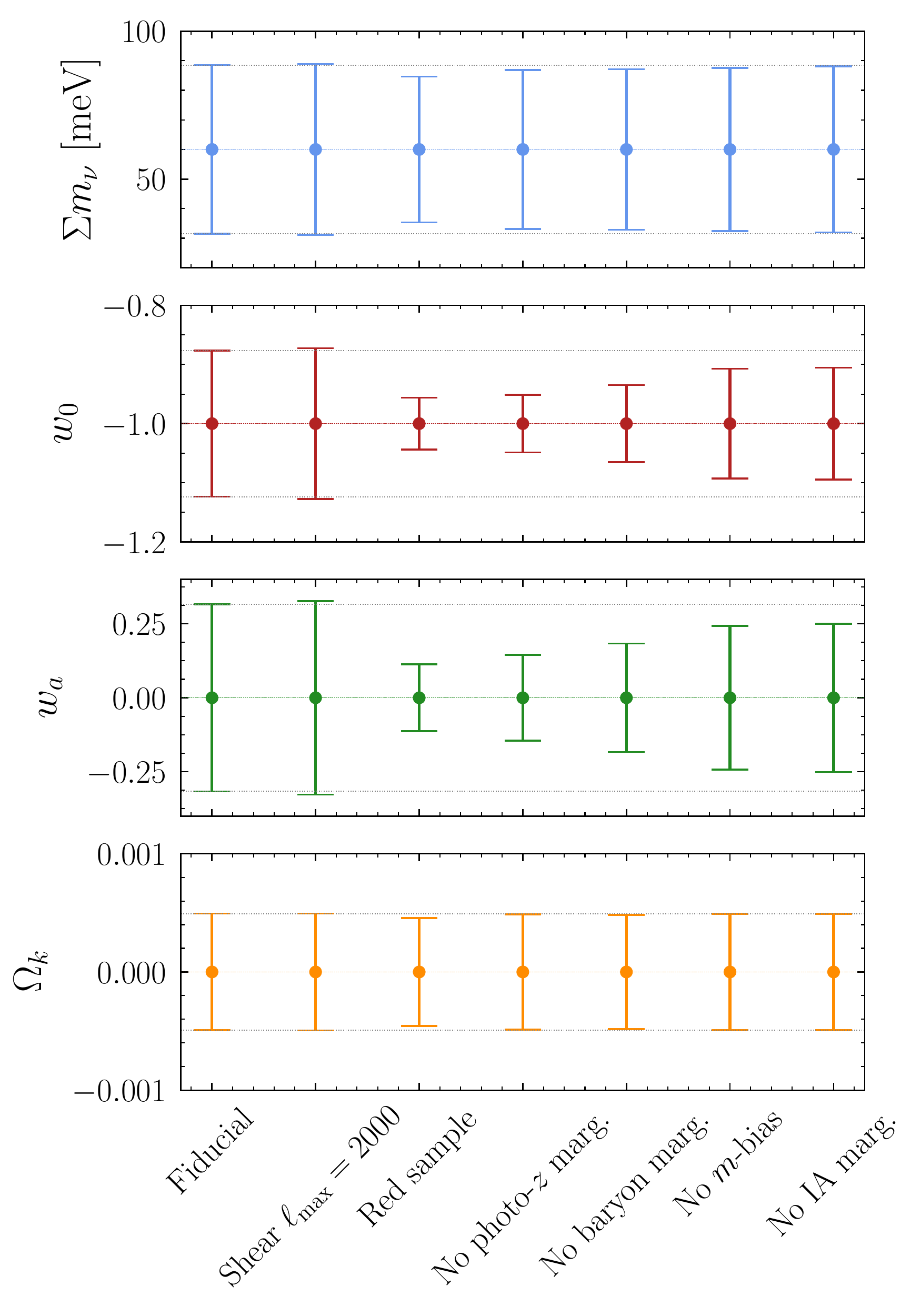}
\caption{Effect of selected systematics on the forecasted neutrino mass sum, dark energy parameters and curvature errors. Each point corresponds to the result of assuming absolute control about each of the systematics, while marginalizing over the rest.}
\label{fig:systematics}
\end{figure}

Ultimately, given their sensitivity, Stage-IV surveys will be limited by systematic effects. Understanding the impact of these systematics is therefore crucial to identify the most critical ones and prioritize their modeling and calibration. In this section we explore the effect of a set of these systematics in turn, as well as the dependence of our final results on the choice of specifications we have made for LSST. The impact of systematics is considered on top of our fiducial setup as described in Sec.~\ref{subsec:mnucomplementarity}, including CMB-S4, LSST shear and clustering along with \emph{Planck} primary CMB. Our findings are summarized in Fig.~\ref{fig:systematics}, which we will refer to in what follows.

\begin{figure}
  \centering
  \includegraphics[width=0.5\textwidth]{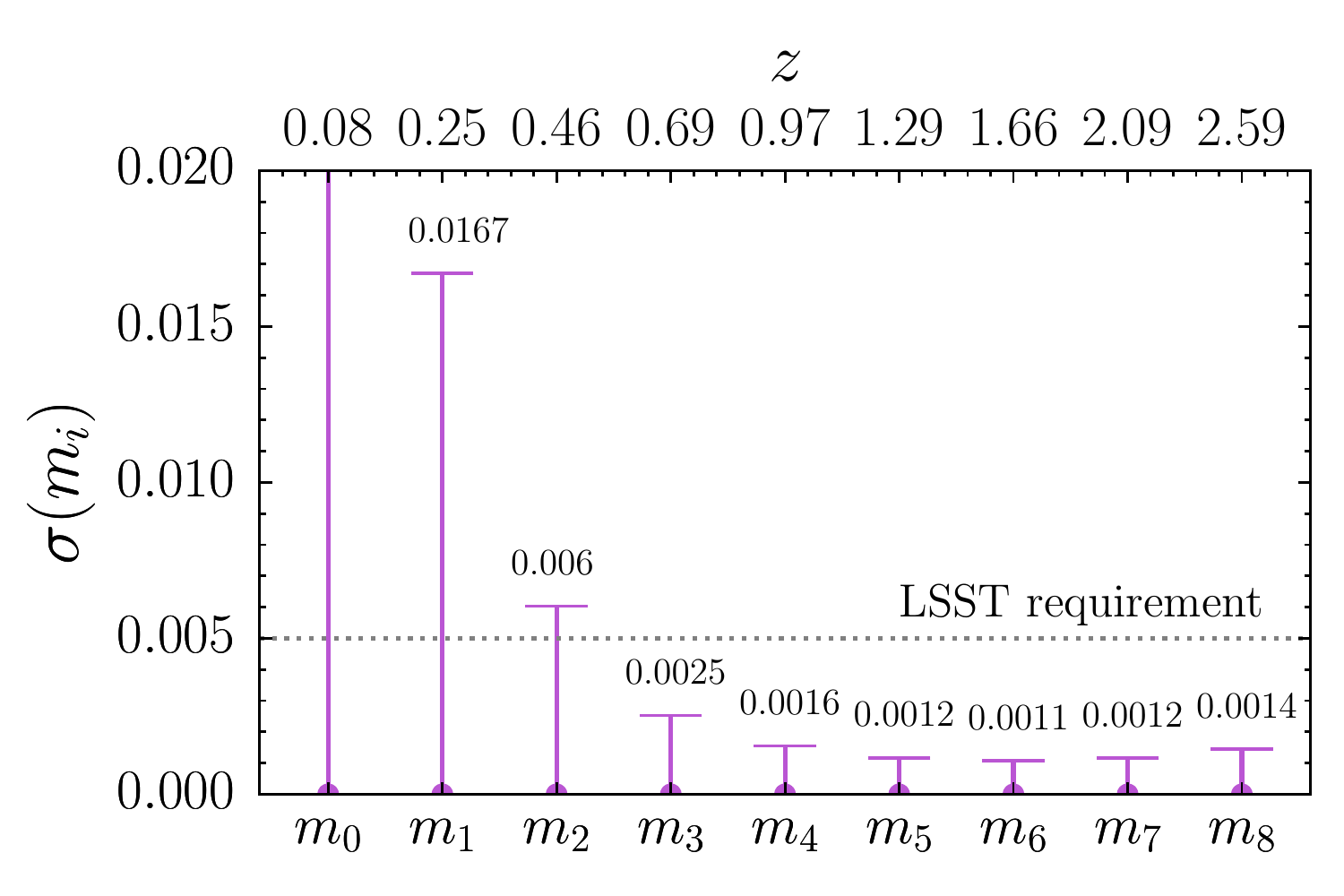}
  \caption{Forecast errors on the multiplicative bias parameters for the different shear bins with LSST and CMB-S4. The dotted line shows the LSST requirement of 0.5\%~\citep{Huterer:2005ez,Massey:2012cz,Schaan:2016ois}.} \label{fig:multibias}
\end{figure}

\begin{itemize}
\item {\sl Scale cuts:} in the baseline case we include shear power spectra up to $\ell_\text{max} = 5000$. Here we show forecasts including only scales up to $\ell_\text{max} = 2000$. This has a small effect on the $\Sigma m_\nu$ forecasts, and results in a slight degradation of the dark energy parameter predictions. This is understandable since, although small scales are much less affected by cosmic variance, weak lensing measurements from LSST become strongly dominated by noise beyond $\ell\sim1000$. This result also shows that the constraints on $\Sigma m_\nu$ derived here are associated mostly with the effect of neutrino masses on the growth of structure, rather than their effect on the shape of the matter power spectrum on small scales.
\item {\sl Multi-tracer clustering:} including an extra clustering sample made up of lower-redshift red galaxies with higher-quality photo-$z$s can improve the final constraints in a number of ways. These include the possible cosmic-variance cancellation through the multi-tracer effect~\citep{Seljak:2008xr}, the improved coverage of the $k_\parallel$-$k_\perp$ plane afforded by the narrower photo-$z$ distributions and the possible self-calibration of photo-$z$ uncertainties in the more numerous blue sample through their cross-correlation with the red galaxies~\citep{Krause:2017ekm}. As we can see in Fig.~\ref{fig:systematics} , we find that including this sample (`Red sample') results in a significant improvement in our ability to measure $w(a)$ (with a $\sim50$\% improvement in both $w_0$ and $w_a$ errors), in agreement with our assessment of the impact of photo-$z$ uncertainties, described below. We find the impact on neutrino masses to be much milder, with only a $\sim10$\% improvement. This agrees with the results of~\citep{Schmittfull:2017ffw} that neutrino masses only benefit mildly from the presence of multiple tracers.
\item {\sl Photo-$z$ uncertainties:} although photometric redshift surveys cannot obtain precise redshift measurements, their success relies on their ability to trace the growth and geometry of the galaxy and matter distributions over different cosmic times, and therefore systematic uncertainties on the redshift distributions of all redshift bins must be kept under control. This is particularly relevant for the measurements of the distance-redshift relation made with galaxy clustering since, unlike shear, this probe provides a measurement of the clustering pattern at the redshift of the sources (and not integrated over a broad kernel).  Figure~\ref{fig:systematics} shows the effect of fixing the bias and scatter parameters for the photo-$z$ distributions. This reduces the dark energy equation of state parameter uncertainties in particular, due to their impact on the distance-redshift relation. We find that the effect on neutrino masses is negligible however, and therefore photo-$z$ calibration requirements are driven by their impact on the dark energy constraints.
\item {\sl Baryonic effects:} since baryonic effects affect the shape of the matter power spectrum at high-$k$, using the same argument as for the scale cuts, the low signal-to-noise of the high-$\ell$ shear power spectrum, combined with the conservative scale cuts applied to the clustering samples imply that the final constraints on $\sigma(\Sigma m_\nu)$ do not suffer much from the possible degeneracy with baryonic parameters. When we hold the baryonic parameters fixed, we find only a small improvement in the dark energy parameters, and no effect on the neutrino mass sum and curvature. Note however that this result may depend on the choice of parametrization for these baryonic effects, and that a more general treatment (e.g., allowing for a redshift dependence of the associated parameters or a more general functional form able to accommodate results from different hydrodynamical models~\citep{Mead:2015yca,Chisari:2018prw}) could affect this degeneracy.
\item {\sl Multiplicative bias:} the multiplicative bias associated with shape measurement systematics can severely limit the ability of lensing shear to constrain the growth of structure as a function of cosmic time. However, we find that there is no significant degradation of $\sigma(\Sigma m_\nu)$ or curvature due to this systematic effect, as shown in Fig. \ref{fig:systematics} (`no m-bias' holds fixed the bias parameters, while they are varied in the fiducial case), and little degradation for the $w_0$ and $w_a$ dark energy parameters. This agrees with the results of~\citep{Schaan:2016ois} -- CMB lensing is able to calibrate the multiplicative bias well within the LSST requirement, with forecast constraints shown in Fig.~\ref{fig:multibias}, and even in the absence of such calibration, low-redshift parameters only suffer mildly from this systematic~\citep{Huterer:2005ez,Massey:2012cz}.
\item {\sl Intrinsic alignments:} the observed shapes of galaxies can be affected, not only by the distortion in the photon path caused by gravitational lensing, but also by local physical forces that alter their actual shapes in a correlated manner. This contaminates the shear power spectrum by up to a few percent, and is an effect that must be taken into account to avoid significant biases in final cosmological parameters. We model IAs using the non-linear alignment model, in which galaxy shapes are proportional to the local tidal field. Uncertainties in, for instance, the alignment amplitude as a function of redshift and magnitude, can lead to significant degradation in cosmological parameters~\citep{Krause:2015jqa}. Our results show a mild, but significant, degradation for the dark energy equation of state parameters $\{w_0,\,w_a\}$ of up to 20\%, in rough agreement with the results of~\citep{Krause:2015jqa}\footnote{The authors of~\citep{Krause:2015jqa} find a quantitatively larger impact for IAs on dark energy constraints. We have studied the impact of IAs for more complicated models, with higher IA amplitudes and a larger number of free parameters (e.g. sampled more finely in redshift), finding consistent results in all cases. We therefore ascribe the differences with~\citep{Krause:2015jqa} to the different modelling assumptions for IAs used there.}. On the other hand, the impact of IAs on the final uncertainty on $\Sigma m_\nu$ is negligible. This can be understood as the IA contamination to the shear power spectrum being self-calibrated through the cross-correlation with galaxy clustering. In fact, a non-negligible degradation of $\sim20\%$ caused by IAs on $\sigma(\Sigma m_\nu)$ is obtained when considering shear measurements only, in the absence of clustering.
\end{itemize}

Finally, although we have shown that our results do not change significantly when altering our assumptions about the main sources of systematic uncertainty, it is worth reminding the reader that the full extent of these effects can only be fully quantified through more sophisticated forecasts. These should include more detailed modeling of all relevant astrophysical uncertainties, as well as a more robust examination of the likelihood function in the presence of systematics than is achievable through a Fisher forecast. It is also worth mentioning that our forecasts assume a Gaussian covariance matrix for the large-scale structure tracers. Non-Gaussianity caused by the non-linear gravitational collapse couples different modes, effectively reducing the total number of degrees of freedom that can be used to constrain cosmological parameters. Thus, although the effect of the non-Gaussian terms have been quantified to be relatively small~\citep{Barreira:2017fjz,Hahn:2018zja}, the absolute forecast uncertainties presented here should be interpreted with care.

\section{Conclusions}
\label{sec:conclusions}

Future cosmological measurements will significantly improve on our ability to constrain non-minimal cosmological scenarios beyond $\Lambda$CDM. In this paper we have studied the degradation in our ability to constrain the sum of neutrino masses with cosmological data associated with allowing for departures from a perfect cosmological constant or a flat Universe. Besides quantifying this degradation, we have also assessed the degree to which it can be mitigated through the combination of data from future CMB experiments, such as CMB-S4, and low-redshift large-scale structure data from galaxy surveys such as LSST. 

Within $\Lambda$CDM, CMB-S4, together with measurements of the BAO scale from DESI should be able to constrain $\Sigma m_\nu$ to the level of $\sigma(\Sigma m_\nu)\sim27\,{\rm meV}$ within $\Lambda$CDM, enough for a 2$\sigma$ measurement of the minimal neutrino mass sum of $60\,{\rm meV}$. However, this measurement would get degraded to $\sigma(\Sigma m_\nu)\sim72\,{\rm meV}$ when allowing for the dark energy equation-of-state parameters $\{w_0,w_a\}$ to vary freely. Combining S4 with tomographic measurements of the growth of structure from a Stage-IV optical galaxy survey such as LSST can, however, bring back the constraining power to the same level as $\Lambda$CDM-only uncertainties by breaking the degeneracies between dark energy and neutrino parameters.

More importantly none of these two experiments would be able measure the minimal $\Sigma m_\nu$ beyond the $2\sigma$ level independently, and combining their datasets will be necessary to reach this sensitivity. In particular, the noise level achievable by CMB-S4 is important to achieve this goal, with a $\sim40\%$ improvement in $\sigma(\Sigma m_\nu)$ from Stage-3-like noise. This improvement with noise level, however, is contingent on the extended parameter space, and is significantly less important if we limit ourselves to $\Lambda$CDM, as noted by~\citep{Allison:2015qca}. Once S4 and LSST are combined, we do not find the DESI BAO measurements to significantly improve the figure of merit $\Sigma m_\nu$, although they could have a significantly positive impact on the dark energy parameters when combined with some individual datasets.

A better measurement of the optical depth to reionization, $\tau$, is an important limitation that future experiments will face when constraining the neutrino mass, a fact that has been noted before in the literature~\citep{Aghanim:2016yuo,Allison:2015qca,Calabrese:2016eii,DiValentino:2015sam,Vagnozzi:2017ovm,Liu:2015txa} in the context of $\Lambda$CDM. This is still the case when confronting an extended parameter set -- we find that both S4 and LSST would benefit significantly from a cosmic-variance limited measurement of $\tau$, and that this is a necessary requirement to go beyond the $2\sigma$-level measurement of the minimal mass sum.

Although the focus of this paper is the measurement of $\Sigma m_\nu$, the formalism used here also allows us to explore the possible constraints on the dark energy equation of state. We find that, unlike in the case of neutrino masses, the measurement of small-scale fluctuations in the CMB from a Stage-4 experiment do not significantly help constrain $\{w_0,w_a\}$ beyond what would already be achievable with LSST alone. This result is dependent on the severity of some of the systematic effects that LSST will have to confront, some of which have been described in the present work. CMB data could help calibrate some of these systematics (e.g.~\citep{Madhavacheril:2017onh,Schaan:2016ois,Giusarma:2018jei}), and therefore the value of combining both datasets cannot be underestimated for any science case.

Finally, we have considered a fairly comprehensive list of systematic effects in our analysis -- photo-$z$ uncertainties, baryonic effects, intrinsic alignments, scale cuts and shear multiplicative biases. We find that the impact of most of the systematic uncertainties that LSST is sensitive to is only marginal in the joint constraints of CMB-S4 and LSST on $\sigma(\Sigma m_\nu)$. This is due to the self-calibration of several of these systematics, and the conservative scale cuts included in our fiducial analysis. On the other hand, we have also shown that $w$ is significantly more sensitive to some of these systematics, especially in the case of photo-$z$ uncertainties. Although we have tried to make a conservative treatment of these systematic effects, the true impact of these will only be quantified through more elaborate forecasts beyond the Fisher matrix formalism and ideally using simulated data with a level of realism commensurate with the expectation of these two experiments. Our forecasts are therefore limited in this sense, and should be interpreted with care. This should not affect the main qualitative message of this work: an optimal measurement of the sum of neutrino masses that can be safely distinguished from the effects of dark energy and non-zero curvature will require the combination of both CMB and large-scale structure data in the Stage-IV era, and cannot be achieved individually by either probe.

\section*{Acknowledgements}
\vspace{-10pt}
We thank Rupert Allison, Jonathan Blazek, Elisa Chisari, William Coulton, Elisabeth Krause and Emmanuel Schaan for useful conversations. DA acknowledges support from the Science and Technology Facilities Council, the Leverhulme and Beecroft trusts and Christ Church college.  This research made use of the \texttt{Astropy}~\citep{Robitaille:2013mpa} and \texttt{IPython}~\citep{PER-GRA:2007}, software packages.

\bibliographystyle{apsrev}
\bibliography{main}

\end{document}